\documentclass{amsart}

\newtheorem{theorem}{Theorem}   
\newtheorem{lemma}[theorem]{Lemma}
\newtheorem{corollary}[theorem]{Corollary}
\newtheorem{proposition}[theorem]{Proposition}

\theoremstyle{definition}

\theoremstyle{remark}
\newtheorem{remark}[theorem]{Remark}

\newcommand{\cA}{{\mathcal A}}

\newcommand{\cH}{{\mathcal H}}

\newsymbol\boxtimes 1202

\newcommand{\bC}{{\mathbb{C}}}
\newcommand{\bN}{{\mathbb{N}}}

\begin{document}

\begin{center}
\vspace*{15mm}
{\LARGE A note on St{\o}rmer condition for decomposability of positive maps\footnote{
This work was partially supported by grant BW/5400-5-0089-8 and SCALA (IST-2004-015714)}}\\

\vspace{2cm}
\textsc{{\large W{\l}adys{\l}aw A. Majewski}\\
Institute of Theoretical Physics and Astrophysics\\
Gda{\'n}sk University\\
Wita Stwosza~57\\
80-952 Gda{\'n}sk, Poland}\\
\textit{E-mail address:} \texttt{fizwam@univ.gda.pl}
\end{center}

\vspace*{4cm}
\noindent
\textsc{Abstract.}

We present a partial characterization of matrices in $M_n(\cA)^+$ satisfying the St{\o}rmer condition.

\vspace{1cm} {\bf Mathematical Subject Classification}: Primary: 46L55
Secondary: 46L05

\vspace{1cm}

\newpage
In \cite{St}
St{\o}rmer gave the following characterization of decomposable maps:
\begin{theorem}{\rm(\cite{St})}
Let $\phi: \cA \to B(\cH)$ be a positive map.
A map $\phi$ is decomposable if and only if for all $n \in \bN$ whenever $(x_{ij})$ and $(x_{ji})$ belong to $M_n(A)^+$ then $(\phi(x_{ij}))\in M_n(B(\cH))^+$.
\end{theorem}
where $\cA$ is a $C^*$-algebra, $B(\cH)$ is the set of all bounded linear operators on a complex Hilbert space $\cH$, and $M_n(\mathfrak{A})$ stands for $n \times n$ matrices over a subspace $\mathfrak{A}$ of a $C^*$-algebra.
Finally $M_n(\mathfrak{A})^+$ denotes the positive part of $M_n(\mathfrak{A})$. Furthermore, throughout this note, $\cH$ will stand for a finite dimensional Hilbert space.

The aim of this note is to provide some elaboration of the condition: $(x_{ij})$ and $(x_{ji})$ are in $M_n(\cA)^+$.
It is worth pointing out that such elaboration seems to be very useful for quantum computing \cite{MMO} as well as for a better understanding of the structure of positive maps (see Question on page 585 in \cite{choi1} and Corollary 7 in \cite{Osaka}).

We firstly  note that
 the positivity of the matrix $(\phi(x_{ij}))$ (with operator entries!) is equivalent to (cf \cite{Tak})
\begin{equation}
\label{dwa}
\sum_{ij} y^*_i \phi(x_{ij}) y_j \geq 0
\end{equation}
where $\{ y_i\}$ are arbitrary elements of $B(\cH)$.
Furthermore, any positive matrix $(x_{ij})$ can be written as

\begin{equation}
\label{hoho}
(x_{ij}) = \sum_k ((v^{(k)}_{i})^* v^{(k)}_{j}
\end{equation}
Hence, applying  condition (\ref{dwa}) to matrices of the form $(a^*_i a_j)$ with the choice of $x_i$ such that all $x_i = 0$ except for $i_0$ and $j_0$, then changing the numeration in such way that $x_{i_0} = x_1$ and $x_{j_0} = x_2$ we arrive to
study the positivity of the following matrix
\begin{equation}
\label{jeden}
\left(
\begin{array}{cc}
a^*_1a_1 & a^*_1a_2 \\
a^*_2 a_1 & a^*_2 a_2 \\
\end{array}  
\right)  
\geq 0            
\end{equation}                
and its transposition.
On the other hand,  block matrix techniques leads to  necessary and sufficient  conditions for positivity of such matrices. Namely, let $A,B,C$ be $d \times d$ matrices. Then
\begin{lemma} {\rm (see \cite{XZ})} 
\label{XZ}
\begin{equation}
\Big[
\begin{array}{cc}
A & B \\
B^* & C \\
\end{array}  
\Big]  
\geq 0            
\end{equation}                
if and only if $A\geq 0$, $C\geq 0$ and there exists a contraction $W$ such that $B = A^{\frac{1}{2}} W C^{\frac{1}{2}}$.
\end{lemma}
Assume, if necessary, that $a_1$ and $a_2$ have inverses otherwise $a_i^{-1}$ is understood to be generalized inverse of $a_i$. Then, an application of Lemma \ref{XZ} to St{\o}rmer condition leads to the following question:
When $|a_1|^{-1} a^*_2 a_1 |a_2|^{-1}$ is a contraction?
But an operator $T \in B(\cH)$ is a contraction if and only if $||T||\leq 1$ what is equivalent to
$||Tx||^2 \leq ||x||^2$. This can be written as
\begin{equation}
\label{contraction}
(x,T^*Tx) \leq (x,x) 
\end{equation} what is equivalent to
\begin{equation}
\label{contraction2}
T^*T\leq \bf 1
\end{equation}

Consequently, (\ref{contraction2}) and Zhan's lemma \ref{XZ} give (see also \cite{A1} and \cite{Choi})
\begin{equation}
a^*_1a_2|a_1|^{-2} a_2^* a_1 \leq |a_2|^2
\end{equation}
Hence
\begin{equation}
\forall_f \quad (f,a^*_1a_2 (a_1^*a_1)^{-1} a^*_2a_1 f) \leq (f, a_2^* a_2 f)
\end{equation}
So, putting $f = a_1^{-1} g$ one gets
\begin{equation}
\forall_g \quad ||(a^*_1)^{-1} a^*_2 g|| \leq ||a_2 a_1^{-1} g||
\end{equation}
This means hyponormality of operators $(a_2 a_1^{-1})^*$ (cf. \cite{H}, and \cite{Stamp}).
But, as considered operators are defined on finite dimensional Hilbert space, in particular, they are completely continuous.
Therefore, hyponormality of $(a_2 a_1^{-1})^*$ implies normality (see \cite{A}, \cite{B}, and \cite{Stamp}).

Consequently, $a_2 a^{-1}$ is a normal operator.
This means that there is a unitary operator $U$ (equivalently unitary matrix as finite dimensions are assumed) such that
\begin{equation}
U a_2 a_1^{-1} U^* = diag(\lambda_i)
\end{equation}
where $\lambda_i \in \bC$. This can be rewritten as
\begin{equation}
a_2a_1^{-1} = \sum_i \lambda_i Q_i
\end{equation}
where $\lambda_i \in \bC$ and $\{Q_i\}$ is the resolution of identity. Hence,
putting $Q_i \equiv |e_i><e_i|$ where $\{ e_i \}$ is a CONS in the Hilbert space $\cH$ on which operators $\{ a_i \}$ act and  \\
$|f><g|z \equiv (g,z) |f>$, one gets
\begin{equation}
\label{lala}
a_2 = \sum_i \lambda_i |e_i><a^*_1e_i|
\end{equation}

Thus we proved:
\begin{proposition}
For any matrix
$\left(
\begin{array}{cc}
a^*_1a_1 & a^*_1a_2 \\
a^*_2 a_1 & a^*_2 a_2 \\
\end{array}  
\right)$  
satisfying the St{\o}rmer condition, $a_2$ is of the form  (\ref{lala}).
\end{proposition}        

\begin{remark}
Using the Ando-Choi inequality (see \cite{A1}, \cite{Choi}) one gets analogous formula for $a_1$ in terms of $a_2$.
\end{remark}
As a next step we note that (\ref{lala}) and St{\o}rmer condition lead to the following form of the matrix $\left(
\begin{array}{cc}
a^*_1a_1 & a^*_1a_2 \\
a^*_2 a_1 & a^*_2 a_2 \\
\end{array}  
\right)$: 

\begin{equation}
\left(
\begin{array}{cc}
a^*_1a_1 & a^*_1a_2 \\
a^*_2 a_1 & a^*_2 a_2 \\
\end{array}  
\right) = \sum_i \left(
\begin{array}{cc}
1 & \lambda_i \\
\bar{\lambda_i} & |\lambda_i|^2 \\
\end{array}  
\right) \dot 
{\Big(
\begin{array}{cc}
|a_1^*e_i><a_1^*e_i| & 0 \\
0 & |a_1^*e_i><a_1^* e_i| \\
\end{array}  
\Big)}   
\end{equation}

To rewrite the above equality in more compact form, let us denote the norm of the vector $|a_1^*e_i>$ by $\alpha_i$ and the normalized vector $\frac{1}{\alpha_i} |a_1^*e_i>$ by $\varphi_i$. Then

\begin{equation}
\label{baba}
\left(
\begin{array}{cc}
a^*_1a_1 & a^*_1a_2 \\
a^*_2 a_1 & a^*_2 a_2 \\
\end{array}  
\right) = \sum_i \alpha_i^2 \left(
\begin{array}{cc}
1 & \lambda_i \\
\bar{\lambda_i} & |\lambda_i|^2 \\
\end{array}  
\right) \dot 
{\Big(
\begin{array}{cc}
|\varphi_i><\varphi_i| & 0 \\
0 & |\varphi_i><\varphi_i| \\
\end{array}  
\Big)}   
\end{equation}
or symbolically
\begin{equation}
\label{chocho}
\left(
\begin{array}{cc}
a^*_1a_1 & a^*_1a_2 \\
a^*_2 a_1 & a^*_2 a_2 \\
\end{array}  
\right) = \sum_i \alpha_i^2  \cdot \Lambda_i \cdot R_i
\end{equation}

where $\Lambda_i$ are ``matrix'' coefficients while $R_i$ are ``matrix'' projectors (not mutually orthogonal!).
This leads to:
\begin{corollary}
 It is well known {\rm{(}}see \cite{Choi2}, \cite{Ster}, and \cite{Wor}{\rm )} that for positive $\phi: M_k(\bC) \to M_l(\bC)$, there is no room for non-decomposable maps when $k=2$, $l=2,3$ but there exist non-decomposable maps for $k=2$, $l=4,5 ...$ In this case, to select decomposable maps one applies St{\o}rmer's Theorem 1.
In this context it is worth pointing out that (\ref{chocho}) implies ``separability'' for $[a_i^*a_j]$ satisfying the St{\o}rmer condition.  Therefore, it is important to realize that non-triviality of St{\o}rmer condition follows from the fact
that when a positive matrix $[x_{i,j}]$ {\rm (}$= \sum_k ((v^{(k)}_{i})^* v^{(k)}_{j}$ {\rm )}  satisfies the St{\o}rmer condition some of its summand(s)
$(v^{(k)}_{i})^* v^{(k)}_{j}$  may not.
\end{corollary}

We end this note with

\begin{remark} 
\begin{enumerate}
\item{} The proof of Proposition 3 is based on  block matrix techniques (cf Lemma 2). However, it should be emphasize that Zhan's lemma can be generalized to more general ``partitions'' . Namely, in a partition $\Big[
\begin{array}{cc}
A & B \\
B^* & C \\
\end{array}  
\Big],  
$ $A$ can be taken to be a $n \times n$ matrix, $C$ to be $k \times k$ matrix, and $B$ to be $n \times k$ matrix respectively. This follows from the observation that the crucial step in the proof of Lemma 2 is the following equivalence:
$\Big[
\begin{array}{cc}
I & X \\
X^* & X \\
\end{array}  
\Big]  
\geq 0
$ if and only if $||X||\leq 1$. However, this equivalence can be extended to the just mentioned more general partitions. Consequently, Lemma 2 can be generalized too. Therefore,  block matrix techniques can be applied directly to an analysis of positive maps $T: M_n(\bC) \to M_k(\bC)$.
\item{} Formula (\ref{baba}) can serve as a recipe for producing PPT states (see \cite{MMO}) and some non-decomposable maps on matrix algebras.
\end{enumerate}
\end{remark}

\section{ Acknowledgements}
The author is very indebted to M. Marciniak for many fruitful discussions and  E. St{\o}rmer for his encouragement and stimulating influence.

\end{document}